# Influence of autocorrelation on the topology of the climate network

Oded C. Guez[1*], Avi Gozolchiani[2] and Shlomo Havlin[1].

[1] Department of Physics, Bar-Ilan University, Ramat-Gan 52900, Israel.

[2] Department of Solar Energy & Environmental Physics, The Jacob Blaustein Institutes for Desert Research, Ben-Gurion University of the Negev, Sede Boqer campus, 84990 Midreshet Ben-Gurion, Israel

Different definitions of links in climate networks may lead to considerably different network topologies. We construct a network from climate records of surface level atmospheric temperature in different geographical sites around the globe using two commonly used definitions of links. Utilizing detrended fluctuation analysis, shuffled surrogates and separation analysis of maritime and continental records, we find that one of the major influences on the structure of climate networks is due to the auto-correlation in the records, that may introduce spurious links. This may explain why different methods could lead to different climate network topologies.

## I. INTRODUCTION

Climate networks are being applied in recent years as a new toolbox for analyzing variations of climate phenomena such as Pacific Decadal Oscillation (PDO), the North Atlantic Oscillation (NAO), the El Niño/Southern Oscillation (ENSO), the North Pacific Oscillation (NPO) and Rossby waves [1–21]. The nodes of the climate network represent geographical locations. The physical process that govern the dynamics of each node are composed of its intrinsic dynamics and coupling terms that depend on the dynamics of other nodes. The cross-correlation between records in two locations is one of the most commonly used measures for determining the links between climate network nodes [22]. The maximum value of the cross-correlation function might appear with a time-delay [23]. The time delay represents the time it takes the climate in one site to influence another site and determine the direction of the link [14].

The literature includes several other definitions for climate network links, such as ordinal patterns and symbolic analysis [24], nonlinear correlation for point processes (event synchronization [25] and symbolic dynamics and renormalized entropy [26]). The outcome network topologies based on these different definitions may vary considerably. For example, Refs. [1,6] observe a high density of links in the tropics and low density of links in the poles, while [3] observe the opposite (see Fig. 1). The current study is aiming to settle and understand the 'order one' difference between two commonly used different schemes by showing that the main difference is due to the influence of autocorrelations on the analysis of links in the climate network.

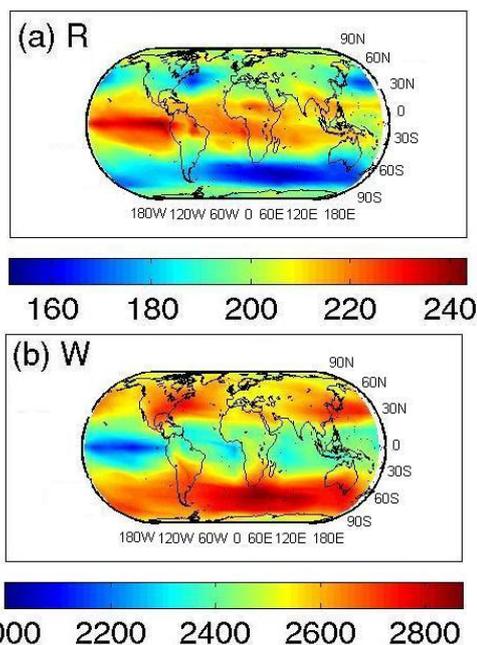

FIG. 1. (Color online) The weighted degree field with weights defined by (a) R defined as the value of the cross correlation peak (see Eq. (2)). (b) W defined as the cross correlation peak, divided by the standard deviation (see Eq. (1)). The weighted degree of a node is defined as the sum of its weighted links [27,28].

* Corresponding author:: oded.guez@biu.ac.il

The paper is organized as follows. The first section is devoted to outlying the methods and databases we used. In the methods section we review two highly used definitions for the calculation of links strength, which in some cases yield different network topologies. We also describe the Detrended Fluctuation Analysis (DFA) method which we will utilize in order to quantify the expected influence of auto-correlation on the analysis of link strength. In the results section we show that the climate network adjacency matrix includes three main sets of links, distinguished in their behavior based on the two different definitions. The first set denoted as A, is where there is an agreement between the two definitions (See Fig. 2a). In the second set denoted as C there exists a contradicting behavior between the two definitions and in the third set denoted as U, both definitions yield links which are in the level of noise and can be disregarded (see Fig. 2a). Afterwards we show, using DFA, that set C is dominated by autocorrelation artifacts. A summary section is then followed.

## II. DATA and METHODS

### A. Data

We analyze the records of the National Center for Environmental Prediction/National Center for Atmospheric Research (NCEP/NCAR) reanalysis air temperature fields at 1000hPa [29]. For each node of the network, daily values for the period 1948-2013 are used, from which we extract anomaly values (actual values minus the climatological averages over the years for each day). We choose 726 nodes covering the globe in an approximately homogeneous manner, so there are 263,175 pairs representing possible links.

To test the role of autocorrelations within the records on the link strength, our control records include shuffled data of two kinds. The first is temperature records with random permutation of full years, where the order of days in each year is maintained. This shuffling preserves all the statistical quantities of the data, such as the distribution of values, and their autocorrelation properties (within one year), but omits the physical dependence between different nodes. The network properties in such a case are only due to the statistical quantities and the autocorrelations of the records and therefore are similar in their properties to spurious links in the original network. To identify unrealistic links, we choose for each link the control records of its pair of nodes and repeat the calculation of the strength of the links. If the strength of the link is significantly higher from that of the control we regard it as real otherwise it is suspected to be a spurious link. The second kind of control is records with random permutation of days. This preserves the distribution of values in the records but omits their autocorrelation properties and the physical dependence between different nodes.

### B. Definitions of a link

Similar to earlier studies [3,11,12], we define the strength of the link measured from data around a date $y$ between nodes $m$ and $n$ as,

$$W_{m,n}^{y} = \frac{MAX(C_{m,n}^{y}) - MEAN(C_{m,n}^{y})}{STD(C_{m,n}^{y})} \qquad (1)$$

where MEAN is the average, STD is the standard deviation and MAX is the maximal value of the absolute value of the cross correlation function $C_{m,n}^{y}$ between the two records. When $W_{m,n}^{y}$ is larger than that of the link of the controlled records we regard it as a real link [11]. We further define the time delay, $\tau_{m,n}^{y}$, as the shift from time zero (time lag in days) corresponding to the time location of highest peak of $C_{m,n}^{y}$. We chose to evaluate Eq. (1) for a sequence of one year and we measure time lags in the range between -219 and +219 days. This interval is chosen to be long enough so that W is not sensitive to our choice of range. The choice of Eq. (1) for identifying significant links is to overcome artificial correlations due to possible persistence or autocorrelations within the records [30].

Another common approach (see e.g. [1,6]) is to define the strength of the link measured from a date $y$ on, connecting the nodes $m$ and $n$, as:

$$R_{m,n}^{y} = MAX(C_{m,n}^{y}). \qquad (2)$$

We compare here between the networks structure obtained using these two definitions, Eqs. (1) and (2).

### C. Detrended Fluctuation Analysis (DFA)

The Detrended Fluctuation Analysis (DFA) [31,32,33] is used for determining how autocorrelation scales with time. We consider a record $\{x_i\}$ of $i = 1,…,N$ equidistant measurements. In most applications, the index $i$ will correspond to the time of the measurements. We are interested in the correlation between the values $x_i$ and $x_{i+s}$ for different time lags, i.e., correlations over different time scales $s$. In order to overcome a constant offset in the data, the mean $\langle x \rangle = 1/N \cdot \sum_{i=1}^{N} x_i$ is usually subtracted, $\overline{x_i} \equiv x_i - \langle x \rangle$. The first step of the algorithm is to transform the time series of the position of a random walker on a 1d chain ('profile') of the record:

$$Y(i) = \sum_{k=1}^{i}(x_k - \langle x \rangle) , \qquad (3)$$

here $i$ can regarded as the first dimension and $Y(i)$ as the second dimension of the walker. The subtraction of the mean $\langle x \rangle$ is not compulsory, since it would be eliminated by the later detrending in the third step of the DFA algorithm. In the second step, we divide the profile $Y(i)$ into $N_s \equiv N/s$ non-overlapping segments of equal length $s$, along the $i$-axis. Since the record length $N$ needs not be a multiple of the considered time scale $s$, a short part at the end of the profile will remain in most cases. In order not to disregard this part of the record, the same procedure is repeated starting from the other end of the record. Thus, $2N_s$ segments are obtained altogether. In the third step, we calculate the local trend for each segment $v$ by a least-squares fit of the profile. Then we define the detrended time series for profile segment of duration $s$, denoted by $Y_s(i)$, as the difference between the original time series and the fits:

$$Y_s(i) = Y(i) - p_v(i). \qquad (4)$$

where $p_v(i)$ is the fitting polynomial in the $v$-th segment. In the fourth step, we calculate - for each of the $2N_s$ segments - the variance:

$$F_s^2(v) = \langle Y_s^2(i) \rangle = \frac{1}{s} \cdot \sum_{i=1}^{s} Y_s^2[(v-1) \cdot s + i], \qquad (5)$$

of the detrended time series $Y_s(i)$ by averaging over all data points $i$ in the $v$-th segment. Finally, we average over all segments and take the square root to obtain the DFA fluctuation function:

$$F(s) = \frac{1}{2N_s} \cdot \sqrt{\sum_{v=1}^{2N_s} F_s^2(v)} . \qquad (6)$$

It is apparent that the variance will increase with increasing duration $s$ of the segments. If the data $(x_i)$ are long-range power-law correlated, the fluctuation functions $F(s)$ increase by a power-law:

$$F(s) \sim s^\alpha. \qquad (7)$$

Here, $\alpha = 0.5$ indicates that there are no correlations. $\alpha > 0.5$ indicates that the data is long-term correlated (the higher $\alpha$, the stronger the correlations are), and $\alpha > 1$ indicates the existence of non stationarities throughout the data. The case $\alpha < 0.5$ corresponds to long-term anticorrelations, large values of $x_i$ are more likely to be followed by small values and vice versa.

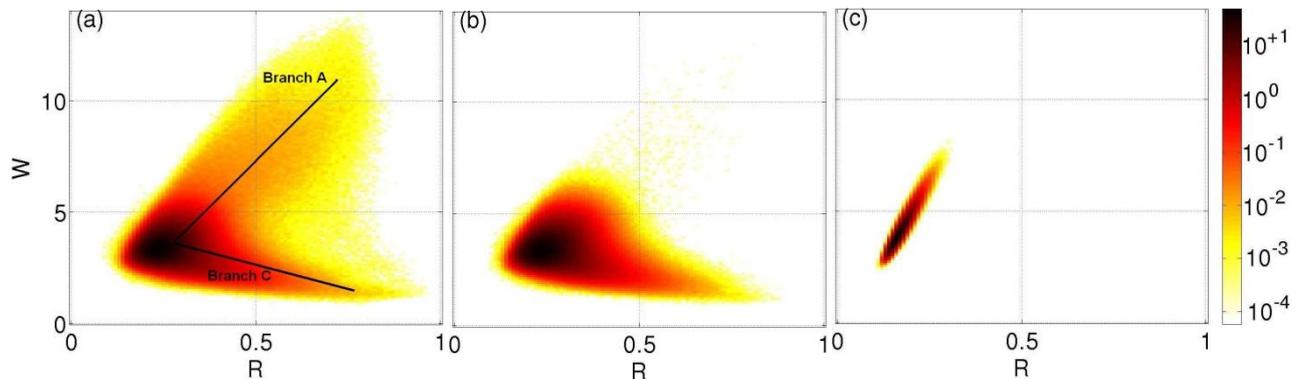

FIG. 2. (Color online) Probability density function of W vs. R (density values are presented in the color bar in the right) for (a) Real data. (b) Records with randomly permuted years (where the order of days in each year is kept). (c) Records with randomly permuted days.

*Corresponding author:: `oded.guez@biu.ac.il`

## III. RESULTS

We show in Fig. 2(a) the probability density function of R and W values for all links of the network. One can see that this 2d distribution has two main branches in which the cross correlation function peak exhibits high value, $R > 0.4$. Branch A defined by $W > 4$ and $R > 0.4$ (representing agreement between W and R) since larger $R$ values correspond to larger W values. Branch C defined by $W > 4$ and $R < 0.3$ representing contradiction between W and $R$ since links with larger $R$ values tend to correspond to smaller W. Additionally there is a third regime U, defined by $R < 0.3$, where the two measures are unrelated (neither correlated nor anti-correlated), which can be regarded as noise. The total number of links in Fig. 2, 17,106,395, is the number of pairs in the network, 263,175 pairs (see Sec. IIA), multiplied by the number of snapshots of the network over all the 63 years. The fraction of links in each category is shown in Table I:

| Branch | A | C | U |
|---|---|---|---|
| Real data | 0.0106 | 0.0248 | 0.7468 |
| Records with permuted years | 0.0027 | 0.0225 | 0.7649 |

TABLE I. Fraction of links in sets A ($R > 0.4, W > 4$), C ($R > 0.4, W < 3$) and U ($R < 0.3$) shown in Fig. 2(a) and (b).

Since we assume that real links have large $R > 0.4$ or large $W > 4$ we focus only on the branches A and C [11]. Note that although the fraction of untrusted noisy links is very high (about 96%) [11], it is found that the trusted links are real and one can learn from their network evolution about climate dynamics [1,3,4,9,14,17,19]. About 20% of the links do not belong to these categories (A, C or U) because of their intermediate nature ($3 < W < 4$), and can be also regarded as unrealistic links.

To investigate the physical properties of links of the contradicting branch C and those in the agreement branch A, we first apply the year-shuffling control (see Sec. IIA) and show in Fig. 2(b) the probability density for pairs W and R of this control analysis. The fraction of links in each category is shown in Table I. Surprisingly, while branch A of Fig. 2(a) almost completely vanishes in Fig. 2(b), branch C almost fully remains, see also Table I. Since the year shuffled pairs of records are not coupled by physical processes, this result may indicate that many non realistic links appear in branch C. In particular, since we preserve the autocorrelation properties in this shuffling scheme, our results suggests that branch C is mostly formed due to autocorrelations as we indeed shown later. Thus, Fig. 2b suggests that mainly large W (i.e., branch A which have also large R) corresponds to realistic climate links.

Next we apply the day-shuffling technique (see Sec. IIA). The network properties in such a case are solely due to the statistical quantities, such as distribution of $x_i$, and therefore are similar in their properties to spurious random links in the original network. Indeed, both branches A and C do not appear in this control test, as expected. The completely random nature of each pair of records in this case results in a correlation function for which each point is a sum of random numbers and hence, due to the central limit theorem, it is distributed normally. The highest peak in each of these correlation functions is distributed according to the generalized extreme value distribution, and therefore implies an almost 1 to 1 relation between W and R. This is evident from the very high correlation (Pearson coeff. = 0.94) found between the two measures, shown in Fig. 2(c).

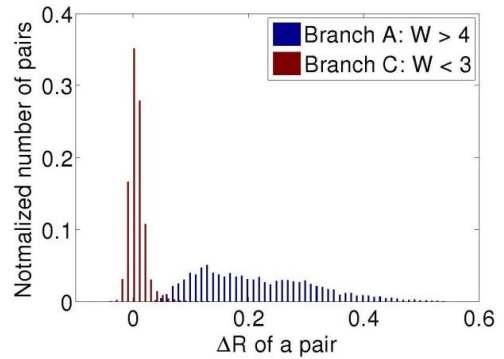

FIG. 3. (Color online) Histogram of the difference between R using real data and R using shuffled data for each link, for branch A (blue) and branch C (red).

Next, we further quantify the effect of yearly shuffling on branches A and C. When a link is formed by a physical coupling process, its weight should be larger when measured in the real system and lower when measured in the control (shuffled years) system. The difference in the R values (defined as $\Delta R$) of each link between the real data and control data is shown in Fig. 3. One can see that for branch C the histogram of links is almost symmetric around 0 (average = 0.01) while for branch A the histogram of links is highly positive (average = 0.22). Since the shuffling does not influence branch C but strongly influence branch A, it further indicates that many unrealistic links appear in branch C.

* Corresponding author:: oded.guez@biu.ac.il

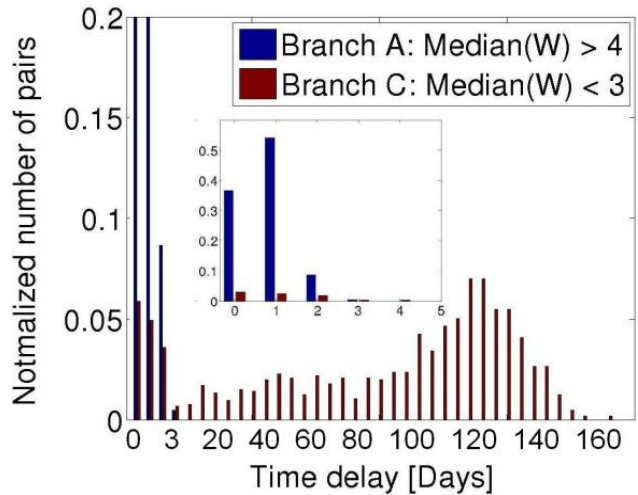

FIG. 4. (Color online) Histogram of the median of the time-delay for branch A (blue) and branch C (red). Note that the fraction of pairs of branch A in the first bin (time delay=0) is 0.36 and in the seconed bin (time delay=1) is 0.52, see Inset showing the first 6 bins (time delay=0-5).

Another test for the reliability of a link is its time-delay, $\tau$. If the time delay is of order of few days (corresponding to weather systerms)we expect the link to be real. However, if the time delay is of order of 100 days it can be considered as an unrealistic link. In Fig. 4 we show the histogram of the median of the time delay of the links over the 66 years in the records. One can see that links of branch A have short time delay (lower than 5 days) while most of the links of branch C have a very broad time delays distribution which peaks around the mid point of the range of time delays we allow.

To understand and further clarify the origin of the different behavior between branch A and C, we show in Fig. 5 an example of a typical link from branch A (up) and a typical link from branch C (down). Most of the links being tested have a similar qualitative shape as the examples shown here. We show in panels (a) and (d) the cross-correlation functions of a typical link of branch A and a typical link of branch C respectively. Links from branch A usually show a  sharp peak with a low background level, while links from branch C show a very slow decay of the peak (with time scales of typically more than 100 days). The differences in the behavior of (a) and (c) can explain the observed differences between branch A and C. When the cross correlation decay slowly, although R might be high, W becomes small since $STD(C_{m,n}^y)$ of Eq.1 is large. In panels (b) and (e) we show typical examples of the time series of each one of the two sites creating the link. The pairs of nodes that correspond to links from branch A seem to have no long-term presistence (or weak one), while links from branch C typically show a clear long term presistence. In panels (c) and (f) we show the autocorrelation function of each one of the two nodes creating the link. Again, links of branch C seem to be due to time series with very fast decaying autocorrelation function, while links of branch A seem to have an autocorrelation function which decays slowly for more than 100 days. The relation between autocorrelations and cross correlations was analyzed for artificial records by Podobnik et al [30] with similar conclusions regarding their effect on the cross correlations. Here we observe this phenomena in real climate records.

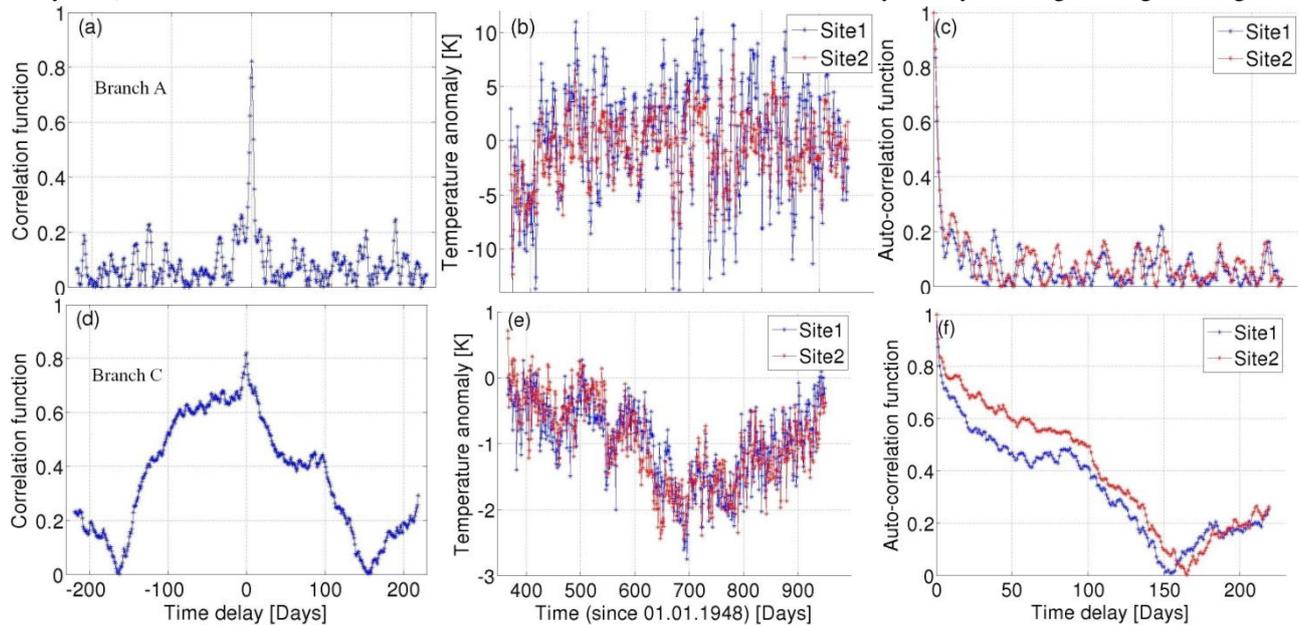

FIG. 5. (Color online) (a) and (d) Typical cross-correlation functions for links in branches A and C respectively. (b) and (e) The time series analyzed in (a) and (d). (c) and (f) The auto-correlation functions of the time series in (b) and (e). The pair of nodes analyzed in (a), (b) and (c) from branch A is $[(22.5N, 15E), (22.5N, 22.5E)]$. The pair of nodes analyzed in (d), (e) and (f) from branch C is $[(7.5S, 100W), (7.5S, 92.5W)]$.

Next we analyze the relation between the auto correlations which is quantified by $\alpha$ (see Sect. II C) and branches A and C in the climate network. We find that pairs with higher values of $\alpha$ are mostly in branch C while those with lower $\alpha$ are in branch A. Fig. 6 depicts $Mean(R)$ vs. $Mean(\alpha)$ where the averages are over different time snapshots of one year. We observe that branch A ($W > 4$) is associated with smaller $\alpha$ compared to branch C ($W < 4$). Moreover, the correlation between $Mean(R)$ and $Mean(\alpha)$ is clearly higher for $Mean(W) < 3$ (Pearson coeff. = 0.61) than for $Mean(W) > 4$ (Pearson coeff. = 0.02). Thus, the association between the link weight R and the autocorrelation is strong for the set C.

The auto-correlation $\alpha$ of nodes inside continent nodes is lower than ocean nodes $\alpha$ due to lower effective heat capacity [34,35]. Fig. 7 shows the probability density function of W vs. R between pairs of continent nodes, pairs of ocean nodes and pairs of mixed ocean and continent nodes. As expected, the ratio between the number of links in branch C and in branch A (see also Table II) is the lowest for links of nodes in continent and highest for links of nodes in ocean.

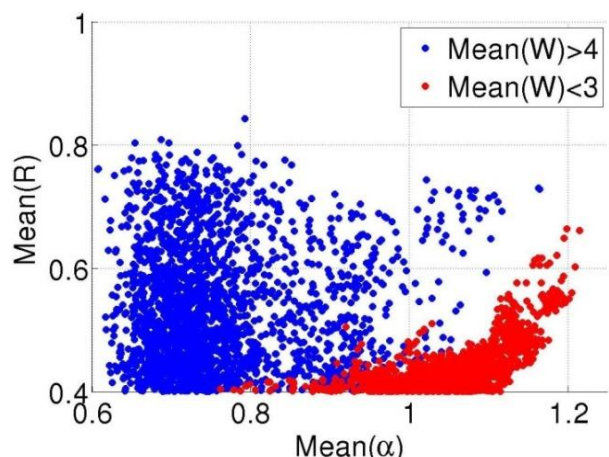

FIG. 6. (Color online) Scatter plot of $Mean(\alpha)$ vs. $Mean(R)$ for $Mean(W) < 3$ and $Mean(W) > 4$ .

| Links in Fig. 7 | Branch C / Branch A |
|---|---|
| (a) Continent | 0.24 |
| (b) Continent and ocean | 2.42 |
| (c) Ocean | 3.27 |

TABLE II. Ratio between the number of links in category A ($R > 0.4, W > 4$) and C ($R > 0.4, W < 3$) for (a) Links between nodes in continent. (b) Links between a node in continent and a node in ocean. (c) Links between nodes in ocean.

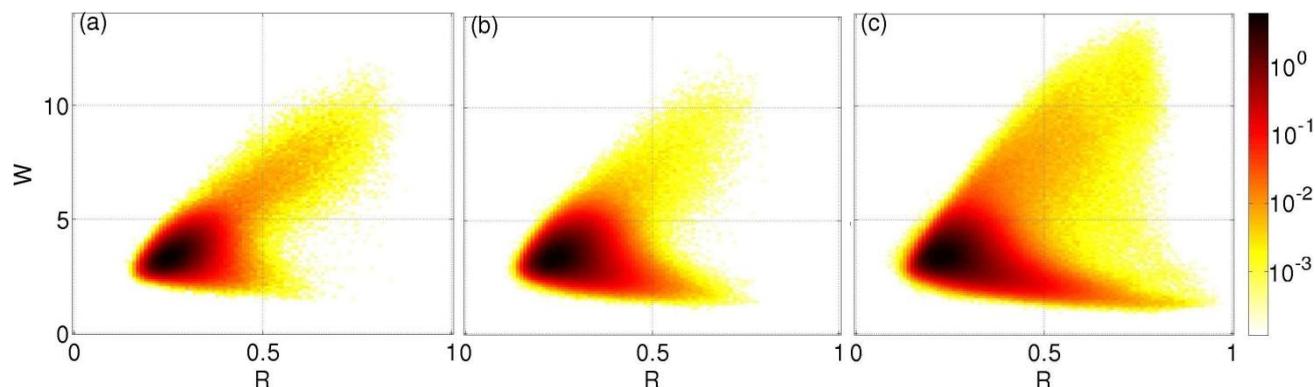

FIG. 7. (Color online) Probability density function of W vs. R for (a) Links between nodes in continent. (b) Links between a node in continent and a node in ocean. (c) Links between nodes in ocean.

* Corresponding author:: oded.guez@biu.ac.il

## IV. SUMMARY

In the current study we compared two commonly used methods for defining links in climate networks. Our analysis is based on constructing two climate networks each using a different method, of atmospheric temperature climate records measured at the surface level on different geographical sites around the globe. Each of the two networks might potentially have groups of links where the two measures correlate, groups of links where the two measures anti-correlate and another group where they are un-related. While correlation is expected (since the two methods are trying to measure the same phenomenon), and unrelated behavior can be attributed to statistically insignificant coupling (and hence random spread of values), the anti-correlated behavior, if exists, is not trivial and is likely to pinpoint systematic (rather than statistical) non-realistic links in one or both methods. We find that the primary reasons for contradicting results between the 2 methods is the existence of significant autocorrelations in the records mostly in ocean, which should be considered when generating climate networks.

## ACKNOWLEDGMENT

We wish to thank the LINC and MULTIPLEX EU projects, DTRA, ONR, the DFG and the Israel Science Foundation for support.

* Corresponding author:: oded.guez@biu.ac.il